\documentclass[%
 aip,
 jmp,%
 amsmath,amssymb,
 reprint,%
]{revtex4-1}

\usepackage{graphicx}
\usepackage{dcolumn}
\usepackage{bm}
\usepackage{braket} 
\usepackage{amssymb,amsmath}
\usepackage{multirow}
\usepackage{epsfig}
\usepackage{hyperref}
\usepackage{subfigure}
\usepackage{bbm}
\usepackage{amsmath}
\usepackage{stmaryrd}
\usepackage{verbatim}
\usepackage{pspicture}
\usepackage{natbib}
\usepackage{verbatim}
\usepackage{tikz}
\usetikzlibrary{decorations.pathmorphing}
\usetikzlibrary{decorations.markings}
\usepackage[english]{babel}
\usepackage{overpic}

\usetikzlibrary{arrows}

\begin{document}
\preprint{AIP/123-QED}

\title[Photon assisted long-range tunneling]{Photon assisted long-range tunneling}
%
\author{Fernando Gallego-Marcos}
  \affiliation{Instituto de Ciencia de Materiales, CSIC, Cantoblanco, 28049 Madrid, Spain}
\author{Rafael S\'anchez}
  \affiliation{Instituto de Ciencia de Materiales, CSIC, Cantoblanco, 28049 Madrid, Spain}
\author{Gloria Platero}
  \affiliation{Instituto de Ciencia de Materiales, CSIC, Cantoblanco, 28049 Madrid, Spain}
\date{\today}

\begin{abstract}
We analyze  long-range transport through an ac driven  triple quantum dot with one single electron. An effective model is proposed for the analysis of photoassisted cotunnel in order to account for the virtual processes which dominate the 
long-range transport, which takes place at n-photon resonances between the edge quantum dots. The AC field renormalizes the inter dot hopping, modifying the levels hybridization. It results in a non trivial behavior of the current  with the frequency and intensity of the external ac field. 
\end{abstract}

\maketitle

%
%
%
\section{Introduction}\label{S:Introduction}
%
%

\begin{figure}[b]
\includegraphics[width=\linewidth]{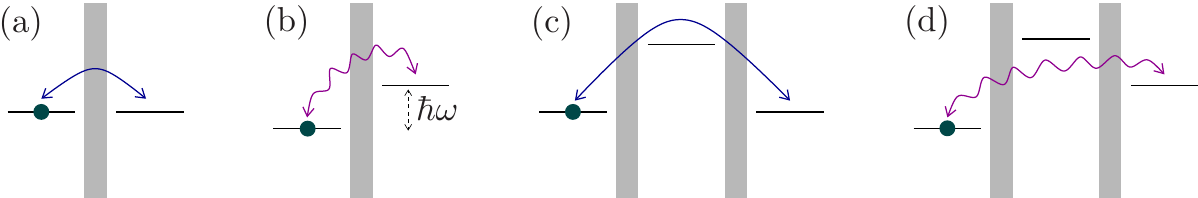}\\
\includegraphics[width=\linewidth]{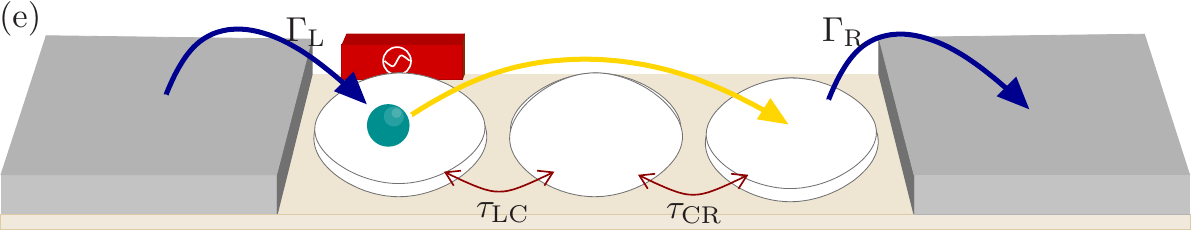}
\caption{(a)-(d) Different tunneling processes: (a) Resonant Rabi oscillations, (b) photon assisted tunneling, $($c) long-range tunneling, (d) photon assisted long-range tunneling. (e) Scheme of a TQD coupled in series to a source and drain. We consider infinite bias between the reservoirs where charge transport is unidirectional from left to right. A time dependent voltage is applied to the left quantum dot only. Long-range tunneling transitions will be induced by the interaction with the ac signal enabling resonant transport in strongly detuned configurations. }
\label{Fig:esquema}
\end{figure}

Quantum transitions between two coupled systems with discrete levels occur when the two levels have the same energy. This is naturally the case in molecules and solids where coherent oscillations are induced. In quantum dots these oscillations can be manipulated by the tuning of both level energies and tunnel couplings with gate voltages~\cite{vanderwiel}. In a double quantum dot, an electron is delocalized between the two dots giving rise to molecular-like orbitals.
When the levels are not resonant, transitions can be induced by the interaction with time dependent fields. The required energy is provided by the absorption or emission of photons with the appropriate frequency, as sketched in Fig.~\ref{Fig:esquema}(b). In transport, this effect is known as photon-assisted tunneling~\cite{report,grifoni} and is accomplished by applying time dependent gate voltages. Remarkably, it allowed for the demonstration of coherent tunneling in semiconductor double quantum dots~\cite{oosterkamp}. Photon assisted transport in nanostructures has been object of intense research~\cite{report}. We mention only those references close to this work in double barriers~\cite{johansson,jesus,ramon,flensberg,pedersen} and quantum dots~\cite{kouwenhoven,oosterkamp,bruder,stoof,stafford,brune,hazelzet,renzoni,strass,spinpump,roman,braakmanPAT,forster}. An additional and important effect of photon assisted tunneling is the renormalization of the tunnel couplings which become
  dependent on the parameters of the driving. This way, a fine tuning of the different couplings, and therefore the control of the hybridized superpositions, is possible by simply acting on the amplitude of the oscillations.

Tunneling between distant states that are not directly coupled (e.g. like next to nearest neighbours in a chain) is also possible even if transitions into the intermediate sites are energetically forbidden. The delocalization in that case is due to higher order transitions~\cite{cohen} when the initial and final states are resonant, with the virtual only occupation of the intermediates, cf. Fig.~\ref{Fig:esquema}$($c). Such a long-range charge transfer is relevant for chemical reactions~\cite{ratner} and has analogues in quantum optics~\cite{cohen}. The recent advances in the tunability of triple quantum dots~\cite{rogge,schroer,gaudreau,amaha, takakura,bischoff,kim} has remarkably achieved the observation of this effect~\cite{bsb,braakman,spinLR}. It entails the transfer of charge or spin between the two ends of the triple dot via superpositions that do not include the center one~\cite{superexchange}, opening the way to the transfer of quantum information between distant qubits with
  low decoherence. In transport through quantum dots, long-range coupling has found interest for the prediction of the interference of spin superexchange trajectories~\cite{superexchange}, the generation of entangled currents~\cite{saraga}, or the influence of dephasing~\cite{debora}. It is also involved in the occurrence of dark states~\cite{michaelis,clive,maria,maria1,andrea}.

In this paper we investigate the transport via the long-range coupling of the outmost sites of a serial triple quantum dot mediated by the interaction with a time dependent potential. This configuration allows for the characterization of single-electron virtual transitions involving discrete levels only. Note the difference with cotunneling events where electrons are transferred from or into a continuum~\cite{averin,ramon,flensberg}. In particular, we consider a configuration where the time dependent signal is applied to one extreme of the array (say the left dot) with the opposite one staying static, cf. Fig.~\ref{Fig:esquema}$($e). We discuss how, surprisingly, an electron in the right dot being static and energetically detuned is coupled resonantly with the left dot via a non-local interaction with the driving. Furthermore, the renormalization of the hopping implies the control of the hybridization which is visible in the splitting of the transport resonance. As a consequence of t
 he driving, St\"uckelberg interference patterns arise for the multiphotonic long-range excitations~\cite{braakman}. As triple dots constitute a tunable three level system, they are ideal platforms to investigate multilevel quantum interference~\cite{forster,danon}.

The paper is organized as follows. In Sec.~\ref{S:TheoreticalFramework} we describe the model and the equations of motion of our triple quantum dot. In Sec.~\ref{S:Transport} the transport characteristics are discussed, and an analytical model is detailed in Sec.~\ref{S:LongRange}. A summary is provided in Sec.~\ref{S:Summary}.

%
%
\section{Theoretical Framework}\label{S:TheoreticalFramework}
%
%
Our system consists on three quantum dots coupled in series, which are denoted with indices L, C and R, the extremes of which are connected to fermionic reservoirs, as sketched in Fig.~\ref{Fig:esquema}. We consider the Coulomb blockade regime with strong electron-electron interactions such that the system can be occupied by up to one electron at a time. Hence we can neglect the effect of spin. The state of the system is then represented by $\{\ket{0},\ket{\text L},\ket{\text C},\ket{\text R}\}$ depending on whether the system is empty or contains an electron in either of the dots.

We describe the system with the Hamiltonian $\hat{H}(t)=\hat{H}_0+\hat{H}_\text{lead}+\hat{H}_\text{int}+\hat{H}_\text{ac}(t)$ with
\begin{align}
\label{eq:HSystem}
\hat{H}_0=&\hat{H}_\epsilon+\hat{H}_\tau=\sum_i\epsilon_i\hat{c}_{i}^\dag\hat{c}_{i}-\sum_{i}\tau_{i,i+1}\hat{c}_{i}^\dag\hat{c}_{i+1}\\
\hat{H}_\text{lead}=&\sum_{lk}\varepsilon_{lk}\hat{d}^\dag_{lk} \hat{d}_{lk},\quad \hat{H}_\text{int}=\sum_{l,k,i}\gamma_{lk}\hat{d}^\dag_{lk}\hat{c}_i+\text{H.c.}
\end{align}
The L-C  and C-R nearest-neighbor hopping is described by $\tau_{ij}$, which is assumed to be real and whose indices are to be counted modulo 3. We emphasize that there is no direct coupling between dots L and R.  $\epsilon_i$ are the quantum dot onsite energies. They include the effect of the capacitive coupling to external gate voltages. $\hat{c}_i$ and $\hat{d}_{lk}$ are the annihilation operators for electrons in dot $i$ and in lead $l$, respectively. 
This ac signal is introduced by a time dependent gate voltage coupled to the left dot:
\begin{align}
\hat{H}_\text{ac}(t)=\frac{V_\text{ac}}{2}\cos(\omega t)\hat{n}_L\label{eq:HAC},
\end{align}
with $V_\text{ac}$ the amplitude of the field and $\omega$ the frequency. 

In order to treat the time dependence, it is convenient to perform a unitary transformation $\hat{U}(t)=\exp\left[(i/\hbar)\int_{0}^t dt'\hat{H}_\text{ac}(t')\right]$ to the Hamiltonian $\hat{\tilde{H}}(t)=\hat{U}(t)(\hat{H}(t)-i\hbar\partial_t)\hat{U}^\dag(t)$. It leaves the diagonal terms of the hamiltonian unaffected and transfers the time dependence to the coupling terms. Thus, using the Jacobi-Anger expansion, the hopping reads:
\begin{align}
\hat{\tilde{H}}_\tau=\sum_{\nu=-\infty}^\infty J_{\nu}(\alpha)\tau_{LC}\text{e}^{i\nu\omega t}\hat{c}_{L}^\dag\hat{c}_{C}
+\tau_{CR}\hat{c}_{C}^\dag\hat{c}_{R}+\text{H.c.},
\label{eq:RenorHTau}
\end{align}
with $J_\nu(\alpha)$ the Bessel functions of the first kind and order $\nu$ and $\alpha=V_\text{ac}/(2\hbar\omega)$. Note that the tunneling coupling $\tau_{LC}$ becomes renormalized: $\tau_{LC}\rightarrow\tilde\tau_{LC,\nu}=J_\nu(\alpha)\tau_{LC}$, which introduces the possibility to tune the coupling by acting on the parameters of the ac field. It also acquires a time-dependent phase which is responsible for the photon assisted tunneling via the exchange of $\nu$ photons with the field. 
Similarly, $\hat{\tilde H}_\text{int}=\sum_{k,i}[\sum_\nu J_\nu(\alpha)\gamma_{Lk}e^{i\omega t}\hat{d}^\dag_{Lk}\hat{c}_L+\gamma_{R}\hat{d}^\dag_{Rk}\hat{c}_R+\text{H.c.}]$.

In the limit of weak coupling to the leads, the dynamics is described by a quantum master equation for the reduced density matrix \cite{stoof,spincorr} of the quantum dot system, $\rho(t)$
\begin{align}
\dot{\rho}(t)=-\frac{i}{\hbar}\big[\hat{H}_\epsilon+\hat{\tilde H}_{\tau}(t),\rho(t)\big]+\mathcal{L}_\Gamma \rho(t) \label{eq:Mastereq}.
\end{align}
The first term of Eq. (\ref{eq:Mastereq}) is the Liouville equation for the coherent time evolution of the density operator of a closed quantum system. The second term describes the tunnel coupling with the reservoirs. Assuming a Born-Markov secular approximation it reads
\begin{align}
\bra{m}\mathcal{L}_{\Gamma}\rho(t)\ket{n}&=\sum_{k\neq n}(\Gamma_{nk}-\Gamma_{kn})\rho_{mn}\delta_{mn}\\&-\frac{1}{2}\left(\sum_{k\neq m}\Gamma_{km}+\sum_{k\neq n}\Gamma_{kn}\right)\rho_{mn}(1-\delta_{mn}).\nonumber
\end{align}
We focus on the high bias regime, where transport is unidirectional from left to right. Furthermore assuming the wide-band limit, all the sidebands couple equally to the left lead. Thus, photon assisted tunneling has no effect in the tunneling through the contacts, following from the properties of the Bessel functions~\cite{ecmi}: $\sum_\nu J_\nu^2(\alpha)=1$. Then the only non-zero tunneling rates $\Gamma_{mn}$ according to the linear configuration with infinite bias are $\Gamma_\text{R}=\Gamma_{R0}=(2\pi/\hbar)|\gamma_{R}|^2\mathcal{D}_D(\hbar\omega_{0R})$ for an electron tunneling into the right lead, and $\Gamma_\text{L}=\Gamma_{L0}=(2\pi/\hbar)|\gamma_{L}|^2\mathcal{D}_S(\hbar\omega_{L0})$ for an electron tunneling from the left lead, where  $\mathcal{D}_l$ is the density of states of each lead.


The dc current is obtained from the stationary solution $\bar\rho$ of Eq. (\ref{eq:Mastereq}). It is simply given by
\begin{align}
I=e\Gamma_\text{R}\bar\rho_{RR},
\end{align}
with $e$ being the elementary charge.


%
\section{Transport}\label{S:Transport}
%

\begin{figure}[t]
	\centering
	\includegraphics[width=0.48\textwidth]{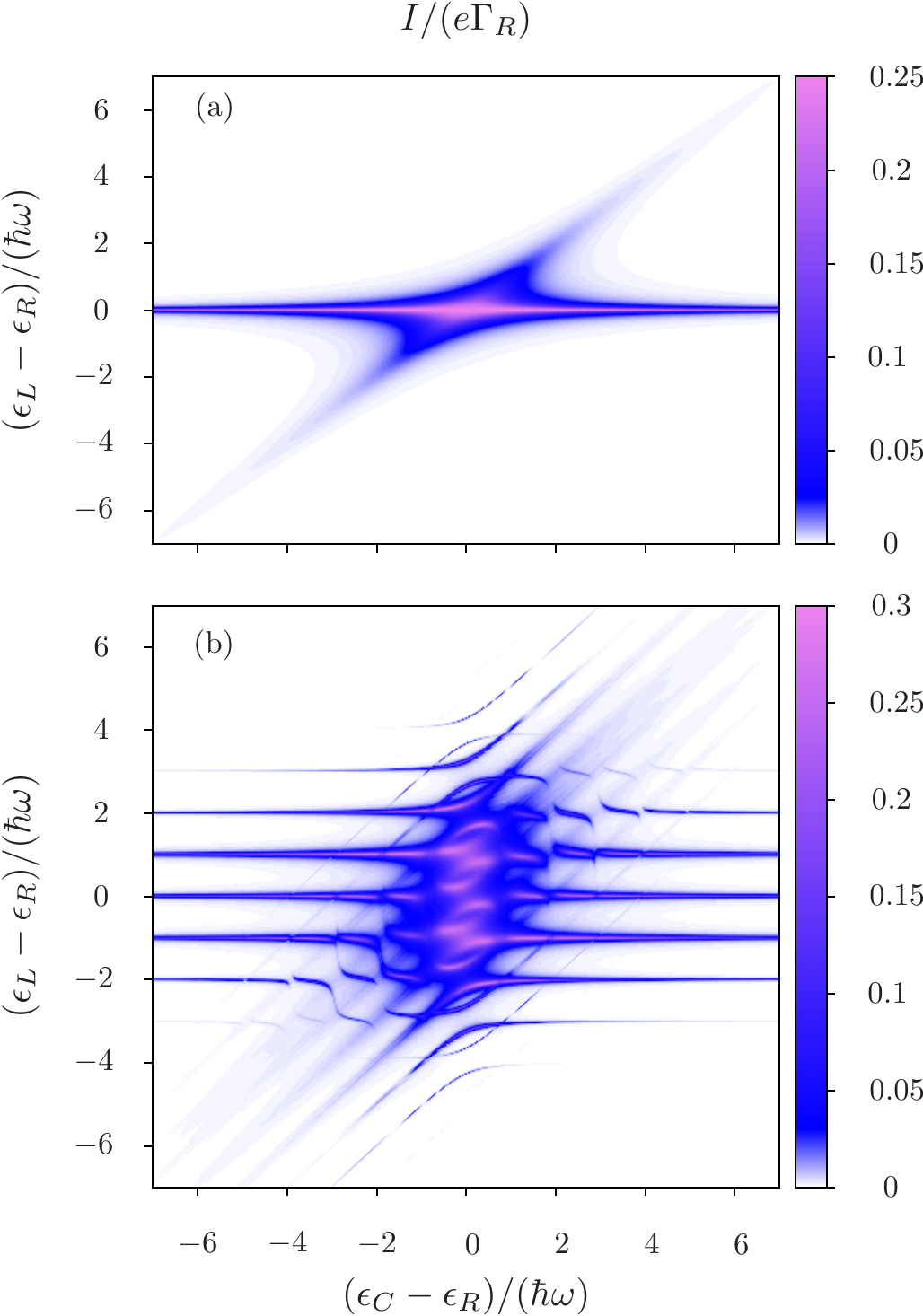}
	\caption{Current through the system $I/e\Gamma_R$ as a function of the level detuning between the energy levels for (a) the undriven and (b) the driven configurations. The amplitude of the oscillations is $V_{ac}/(2\hbar\omega)=1.52$. The coupling between the dots are symmetric $\tau_{LC}=\tau_{CR}=\tau=10\Gamma$, with $\Gamma_\text{L}=\Gamma_\text{R}=\Gamma$. }
	\label{FIG:NOAC}
\end{figure}

Let us first discuss the transport characteristics of the undriven case. Current is expected to flow only close to the quadruple point when the four charge configurations are degenerate. In fact, as shown in Fig.~\ref{FIG:NOAC}(a), a large current peak appears around the condition $\varepsilon_L=\varepsilon_C=\varepsilon_R$ where transport is resonant through all the structure. The current extends along the condition $\varepsilon_L=\varepsilon_C$ but rapidly faints due to the finite but small broadening of the level in the dot coupled to the drain, $\Gamma_R$. Most interestingly for us, a very narrow resonance occurs along the condition $\varepsilon_L=\varepsilon_R$ that survives far from the quadruple point, where the center dot is detuned. This is the long-range transfer that we are interested in. There, the electron is delocalized between the left and right dots with the center one being only virtually occupied. This is clearly represented in the eigenstate $\tau_{CR}|L\rangle-\tau_{LC}|R\rangle$, see Appendix~\ref{A:Undriven} for more details. Such a resonance has been observed experimentally involving the correlated tunneling of two electrons~\cite{bsb} or of a single spin~\cite{spinLR}. Let us mention that when the couplings are inhomogeneous, i.e. $\tau_{LC}\ne\tau_{CR}$, one of the side dots hybridize more strongly with the center one effectively forming two molecular-like orbital (cf. Appendix~\ref{A:Undriven}). The other dot thus crosses two levels which gives rise to a splitting of the central peak~\cite{spinLR}.

We focus now on the driven configuration where an ac potential is applied to the left dot. As discussed above, the effect of the driving is two-fold. On one hand, it induces photon assisted transport in detuned configurations when the frequency of the oscillation matches an integer fraction of the energy difference, $\hbar\omega\approx\nu^{-1}(\varepsilon_i-\varepsilon_j)$. Then, the delocalization takes place via the absorption/emission of $\nu$ photons. As a consequence, the different resonances split into a number of sidebands, as plotted in Fig.~\ref{FIG:NOAC}(b). 

On the other hand, the tunneling couplings get  renormalized~\eqref{eq:RenorHTau} due to the time dependent voltage. This makes the height and width of the different peaks become dependent of the details of driving, in particular of the argument of the Bessel functions, $V_\text{ac}/(2\hbar\omega)$. For the values considered here, sidebands with $\nu>\pm5$ are weak. Also, the renormalization affects the overlapping of the different states. As $J_\nu(\alpha)\tau_{LC}<\tau_{CR}$, the center dot hybridizes more strongly with the right dot, effectively forming an extended two level dot around $\varepsilon_C=\varepsilon_R$. This is visible in the splitting of the central peak around each sideband into some kind of avoided crossings, cf. Fig.~\ref{FIG:NOAC}(b). Consistently the splitting is better resolved  as the order of the sideband increases, which corresponds to a smaller $J_\nu(\alpha)$.

Additional features appear as current minima on the crossing of L-R and L-C resonances involving a different number of photons. In this case, the electron in the left dot can tunnel resonantly either to the center (by direct tunneling) or to the right dot (via long-range tunneling). In the former case, as the subsequent transition  between C and R dots is energetically forbidden, the occupation of the center dot reduces the current.


%
%
\section{Long-Range Tunneling}\label{S:LongRange}
%

\begin{figure}[t]
	\centering
	\includegraphics[width=0.48\textwidth]{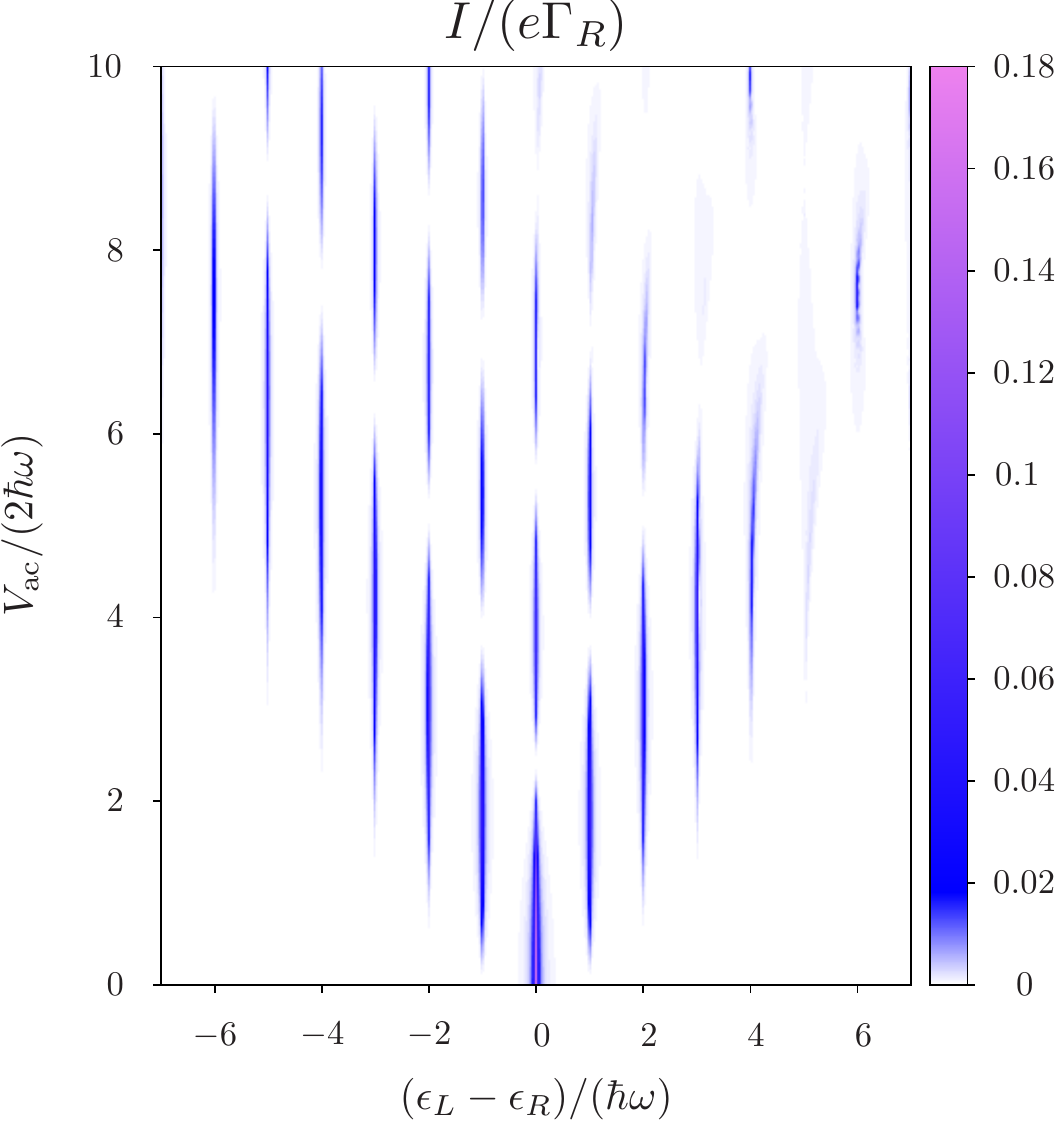}
	\caption{Current through the system as a function of the amplitude of the ac field and the detuning between the L and R dots. The C dot is detuned by $10\hbar\omega$, far away from the L-R resonances. Its effect is only visible as some asymmetry in the high detuning and strong driving regime.}
	\label{FIG:RWAOneVEC}
\end{figure}
Let us focus now on the regime $\varepsilon_R-\varepsilon_L\approx n\hbar\omega$ where long-range tunneling dominates. We have mentioned that the interdot hopping $\tau_{LC}$ is affected by the parameters of the driving through Bessel functions. In order to further understand its effect on the left-right delocalization, we investigate the photon-assisted current as a function of the detuning $\varepsilon_L-\varepsilon_R$ and the amplitude of the oscillations $V_\text{ac}$. The result is plotted in Fig.~\ref{FIG:RWAOneVEC}, where well-defined sidebands are resolved at the conditions $\varepsilon_L-\varepsilon_R\approx n\hbar\omega$. Within each resonance involving $n$ photons, the current becomes strongly dependent on $J_n(\alpha)$, as expected from the renormalization of $\tau_{LC}$. This effect is evident in the points where the current vanishes which coincide with the zeros of the Bessel functions. Interestingly, the number of involved photons is determined solely by the detuning b
 etween the initial and final states of the transition, $|L\rangle$ and $|R\rangle$. The detuning of these states with respect to $|C\rangle$ only affects the strength of the virtual transition, as we show in more detail in the following by performing a perturbative expansion and a rotating way approximation (RWA). 
The pattern in Fig.~\ref{FIG:RWAOneVEC} can also be interpreted as multiphoton Landau-Zener-St\"uckelberg-Majorana tunneling interferometry pattern~\cite{shevchenko} notably involving virtual transitions, which are here resolved in a transport configuration. 



\subsection{Analytical treatment}\label{SS:PerturbationExpansion}

Further understanding of the photon-assisted long-range tunneling close to the condition $\epsilon_L-\epsilon_R+n\hbar\omega=0$ is gained by considering the coherent dynamics in the closed triple quantum dot system. This is given by the Schr\"odinger equation
\begin{align}
[\hat{H}_\epsilon+\hat{\tilde{H}}_\tau(t)]\ket{\Psi_S(t)}=i\hbar\frac{\partial}{\partial t}\ket{\Psi_S(t)}.\label{eq:Spicture}
\end{align}
In the interaction picture, $\ket{\Psi_S(t)}=\exp[-i/\hbar \hat{H}_\epsilon (t-t_0)]\ket{\Psi_I(t)}$, the time evolution of the wave function reads,
\begin{align}
\ket{\Psi_I(t)}&=e^{-\frac{i}{\hbar}\int^{t}_{t_0}\hat{\tilde{H}}_{\tau,I}(t')dt'}\ket{\Psi_I(t_0)}\label{eq:cotI}
\end{align}
We can expand the time evolution operator:
\begin{align}
\label{eq:cotSeries}
\mathcal{U}(t,t_0)=&1-\frac{i}{\hbar}\int^t_{t_0}dt_1\hat{\tilde{H}}_{\tau,I}(t_1)\\
&-\frac{1}{\hbar^2}\int^t_{t_0}dt_1\int^{t_1}_{t_0}dt_2\hat{\tilde{H}}_{\tau,I}(t_1)\hat{\tilde{H}}_{\tau,I}(t_2)+\ldots \nonumber
\end{align}
We are seeking the coupling between $\ket{\text L}$ and $\ket{\text R}$ which is not provided by the two first terms of the expansion.
At second order, we get: 
\begin{widetext}
\begin{align}
\mathcal{U}^{(2)}(t,0)\ket{\text L}
=&\sum_\nu \frac{\tau_{LC}(-1)^{\nu}J_\nu(\alpha)}{\epsilon_L-\epsilon_C+\nu\hbar\omega}\left\{\sum_{\nu'}\tau_{LC}J_{\nu'}(\alpha)\left(\frac{e^{i(\nu+\nu')\omega t}-1}{(\nu+\nu')\hbar\omega}-\frac{e^{i(\epsilon_C-\epsilon_L+\nu'\hbar\omega)t/\hbar}-1}{\epsilon_C-\epsilon_L+\nu'\hbar\omega}\right)\ket{\text L}\right.\nonumber\\
&\quad\left.+\tau_{CR}\left(\frac{e^{i(\epsilon_L-\epsilon_R+\nu\hbar\omega)t/\hbar}-1}{\epsilon_L-\epsilon_R+\nu\hbar\omega}-\frac{e^{i(\epsilon_C-\epsilon_R)t/\hbar}-1}{\epsilon_C-\epsilon_R}\right)\ket{\text R}\right\}\\
\mathcal{U}^{(2)}(t,0)\ket{\text R}
=&-\frac{\tau_{CR}}{\epsilon_R-\epsilon_C}\left\{\sum_{\nu}\tau_{LC}J_{\nu}(\alpha)\left(\frac{e^{-i(\epsilon_L-\epsilon_R-\nu\hbar\omega)t/\hbar}-1}{\epsilon_L-\epsilon_R-\nu\hbar\omega}+\frac{e^{i(\epsilon_C-\epsilon_L+\nu\hbar\omega)t/\hbar}-1}{\epsilon_C-\epsilon_L+\nu\hbar\omega}\right) \ket{L}\right.\nonumber\\
&\left.\quad+\tau_{CR}\left(\frac{t}{i\hbar}-\frac{e^{-i(\epsilon_R-\epsilon_C)t/\hbar}-1}{\epsilon_R-\epsilon_C}\right)\ket{R}\right\},
\end{align}
\end{widetext}
i.e. the left and right dots get coupled via second (and higher) order transitions characterized by the parameters $\tau_{LC}/(\epsilon_C-\epsilon_L-\nu\hbar\omega)$ and $\tau_{CR}/(\epsilon_C-\epsilon_R)$. 
We are interested in the case where the center dot is decoupled from any resonant transition (including those assisted by photons), i.e. $\tau_{LC}\ll\epsilon_C-\epsilon_L-\nu\hbar\omega$ and $\tau_{CR}\ll\epsilon_C-\epsilon_R$, so we can neglect any other contribution to the expansion.
Transforming it back into the Schr\"odinger picture and taking the time derivative in the right-hand side of Eq.~\eqref{eq:Spicture},
we arrive at the effective Hamiltonian:
\begin{align}
\hat{H}_{\text{eff}}
=\left(\begin{array}{cc}
\tilde\epsilon_L &-\sum_\nu(-1)^\nu g_{LR,\nu}e^{i\nu\omega t}\\
-\sum_{\nu} g_{LR,\nu}e^{i\nu\omega t}&\tilde\epsilon_R\\\end{array}\right),
\label{eq:effHS}
\end{align}
with the shifted energies $\tilde\epsilon_L=\epsilon_L-\sum_\nu\sum_{\nu'} g^{\nu\nu'}_{LL}e^{i(\nu+\nu')\omega t}$ and $\tilde\epsilon_R=\epsilon_R-g_{RR}$ and where we have introduced the coefficients $g^{\nu\mu}_{LL}=(-1)^{\nu}\tilde\tau_{LC,\nu}\tilde\tau_{LC,\mu}/(\epsilon_L-\epsilon_C+\nu\hbar\omega)$, $g_{RR}=\tau_{CR}^2/(\epsilon_R-\epsilon_C)$ and 
\begin{equation}
g_{LR,\nu}=\frac{\tilde\tau_{LC,\nu}\tau_{CR}}{\epsilon_L-\epsilon_C+\nu\hbar\omega}.
\end{equation}
We have neglected fast oscillating terms including $\exp{[i\hbar^{-1}(\epsilon_C-\epsilon_R)t]}$ and $\exp{[i\hbar^{-1}(\epsilon_C-\epsilon_L-\hbar\omega)t]}$ which are assumed to be integrated out in the dynamics.
The effective coupling has the expected cotunneling-like form~\cite{flensberg} which contains the coupling due to both barriers (the left one being renormalized with a Bessel function) and decreases as the inverse of the distance to the intermediate virtual state~\cite{superexchange}.

\subsection{Rotating wave approximation}\label{SS:RWA}

The effective Hamiltonian~\eqref{eq:effHS} is equivalent to that of a driven two level system, $\ket{\tilde L}$ and $\ket{\tilde R}$. It describes resonances when the exponent $n\omega$ matches the energy difference $(\tilde\epsilon_R-\tilde\epsilon_L)/\hbar$. A standard technique to investigate the dynamics close to each of these resonances is to apply a rotating wave approximation~\cite{grifoni} that neglects any off-resonant transition. 
Following Ref.\onlinecite{sanchez} we transform 
Eq. (\ref{eq:effHS}) into the rotating  frame, $\hat{H}_{\text{RWA}}=\mathcal{V}^\dag \hat{H} _{\text{eff}}\mathcal{V}$, with the unitary vector
\begin{align}
\mathcal{V}=\hat{c}^\dag_{\tilde L}\hat{c}_{\tilde L}+e^{-in\omega t}\hat{c}^\dag_{\tilde R}\hat{c}_{\tilde R} \label{eq:UU}
\end{align}
that shifts the energy by $n\hbar\omega$. 
With this notation the rotated Hamiltonian reads
\begin{align}
\hat{H}_{\text{RWA}}&=\left(\begin{array}{cc}\tilde{\epsilon}_{L,n}&(-1)^{n-1} g_{LR,n}\\
(-1)^{n-1}g_{LR,n}&\tilde{\epsilon}_R-n\hbar\omega\\\end{array}\right)
\label{eq:HRWA}
\end{align}
with $\tilde{\epsilon}_{L,n}=\epsilon_L-g_{LL}^{nn}$ and $\tilde{\epsilon}_R=\epsilon_R-g_{RR}$ , where we have neglected all the oscillating terms. 
The resulting Hamiltonian being time independent, it can be diagonalized. This way, we find the eigenstates 
\begin{align}
\ket{\pm}=\frac{1}{\sqrt{2}}\left(\ket{\tilde{\text L}}\pm\ket{\tilde{\text R}}\right)
\end{align}
that describe the resonant charge transfer between the left and right dot, analogously to~\eqref{eq:LongEst} and\eqref{eq:doubleDot} in the undriven case. 

When open to transport, the system thus becomes equivalent to a double quantum dot. For that system, the current can be analytically calculated~\cite{gurvitz}, giving a Lorentzian-shaped resonance:
\begin{equation}
\label{eq:IRWA}
I_{\text{RWA}}^{(n)}=\frac{eg_{LR,n}^2\Gamma_\text{L}\Gamma_\text{R}}{g_{LR,n}^2\Gamma_\text{T}+\Gamma_\text{L}[\Gamma_\text{R}^2+4(\Delta-n\hbar\omega)^2]},
\end{equation}
with $\Delta=\tilde\epsilon_R-\tilde\epsilon_L$ and $\Gamma_\text{T}=2\Gamma_\text{L}+\Gamma_\text{R}$.
 
\begin{figure}[t]
	\centering
	\includegraphics[width=0.48\textwidth]{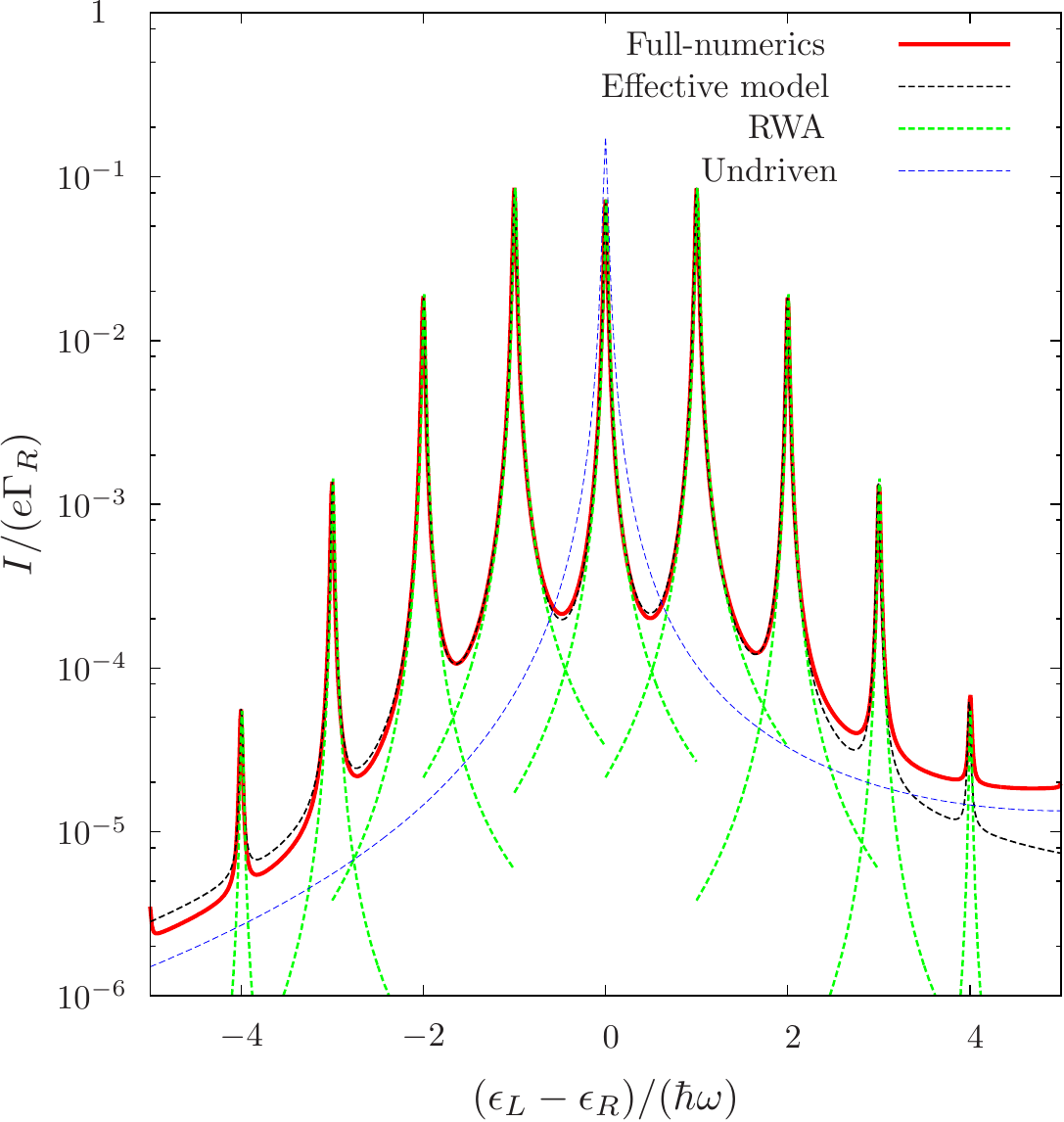}
	\caption{Current as a function of the left-right detuning in the long-range tunneling regime, with $\epsilon_C-\epsilon_R=10\hbar\omega$. We compare the full numerics solution, obtained by the integration of Eq.~\eqref{eq:Mastereq} with the different approximations: the effective model for the tunnel coupling~\eqref{eq:effHS} and the rotating wave approximation~\eqref{eq:IRWA}. We also plot the current for the undriven case which only shows the central peak. The driving is characterized by the parameter $\alpha=V_\text{ac}/(2\hbar\omega)=1.38$. For $\epsilon_L-\epsilon_R>3\hbar\omega$, a deviation appears due to the presence of the resonance with the center dot. }
	\label{FIG:RWAOne}
\end{figure}
In order to test the validity of the different approximations we made, we plot Fig.~\ref{FIG:RWAOne}. There, we compare the current obtained from the numerical integration of the full master equation~\eqref{eq:Mastereq} with that obtained by considering the second-order expansion effective Hamiltonian~\eqref{eq:effHS} and with the analytical expression for the current in the rotating wave approximation~\eqref{eq:IRWA} around each resonance, finding a remarkably good agreement. For completeness, we also plot the current obtained in the undriven configuration, where only the central peak appears, which clearly shows the reduction of the current intensity and the narrowing of the current peak in the presence of  the ac voltage due to the renormalization of $\tau_{LC}$. 
\section{Summary}\label{S:Summary}
%
%

In summary, we analyze photon assisted long range transport through a linearly coupled triple quantum dot. The transport between the edge quantum dots is assisted by the ac field giving rise to resonances between the left and right quantum dot levels differing in $n$ photons. The ac driving also renormalizes the interdot hopping and in consequence the levels hybridization. Therefore the ac field determines the current spectrum and allows its control by tuning the ac field parameters.

We present as well a theoretical model of photon assisted second order transitions (cotunnel) and propose an effective Hamiltonian. We have considered the simplest model with up to one electron. The analysis of other regions in the stability diagram with more electrons will be affected by charge and spin correlations. 




\appendix
%
%
\section{Undriven Case}\label{A:Undriven}
%
%
In this appendix we present in more detail the case without an ac field coupled to the QD system. The Hamiltonian Eq. (\ref{eq:HSystem}) without AC in matrix form reads
\begin{align}
\hat{H}_0=\left(\begin{array}{ccc}\epsilon_L&\tau_{LC}&0\\\tau_{LC}&\epsilon_C&\tau_{CR}\\0&\tau_{CR}&\epsilon_R\\\end{array}\right).\label{eq:H0}
\end{align}
The eigenstates for the condition of the long range coherent state ($\epsilon_L=\epsilon_R=\epsilon$) are
\begin{align}
\ket{\text LR}=&\tau_{CR}\ket{\text L}-\tau_{LC}\ket{\text R}\label{eq:LongEst}\\
\ket{\pm}=&\tau_{LC}\ket{\text L}+\frac{1}{2}(\delta_{C}\pm\delta_\tau)\ket{\text C}{+}\tau_{CR}\ket{\text R},\label{eq:doubleDot}
\end{align}
with $\delta_{C}=\epsilon_C-\epsilon$ and $\delta_\tau=\sqrt{\delta_{C}^2+4(\tau_{LC}^2+\tau_{CR}^2)}$, and the respective eigenenergies $E_\text{LR}=\epsilon$ and $E_{\pm}=(\delta_{C}\pm\delta_{C}{\pm}\delta_\tau)/2$.
If the coupling between the dots are the same, the left and right dots have the same weight in Eq. (\ref{eq:LongEst}) and are equally hybridized with the central dot in Eq. (\ref{eq:doubleDot}). 

In the case asymmetric case $\tau_{i}\gg\tau_{j}$ with $\delta_C\gg\tau_i$, the states of two of the quantum dots ($\ket{\text C}$ and $\ket{j}$) get strongly hybridized and effectively form a two level quantum dot to which the state of the other dot, $\ket{i}$, is coupled. This is reflected in the eigenstates
\begin{align}
\ket{\text LR}\approx&\ket{j},\quad i\neq j \in\{\text L,\text R\}\\
\ket{\pm}\approx&\ket{\text C}\pm\ket{i}.
\end{align}

\acknowledgements
We acknowledge financial support from the Spanish MICINN MAT2011-24331 and Juan de la  Cierva program (RS).


\end{document}